\documentclass[%
 reprint,
 amsmath,amssymb,
 aps,
pra,
]{revtex4-1}

\usepackage{float}
\usepackage{graphicx}
\usepackage{dcolumn}
\usepackage{bm}

\begin{document}

\preprint{APS/123-QED}

\title{Collective modes of a harmonically trapped one-dimensional Bose gas: the effects of finite particle number and nonzero temperature}

\author{Xiao-Long Chen}
\author{Yun Li}
\author{Hui Hu}%
 \email{hhu@swin.edu.au}
\affiliation{%
 Centre for Quantum and Optical Science, Swinburne University of Technology, Melbourne, Victoria 3122, Australia
}%

\date{\today}

\begin{abstract}
Following the idea of the density functional approach, we develop a generalized Bogoliubov theory of an interacting Bose gas confined in a one-dimensional harmonic trap, by using a local chemical potential - calculated with the Lieb-Liniger exact solution - as the exchange energy. At zero temperature, we use the theory to describe collective modes of a finite-particle system in all interaction regimes from the ideal gas limit, to the mean-field Thomas-Fermi regime, and to the strongly interacting Tonks-Girardeau regime. At finite temperature, we investigate the temperature dependence of collective modes in the weak-coupling regime by means of a Hartree-Fock-Bogoliubov theory with Popov approximation. By emphasizing the effects of finite particle number and nonzero temperature on collective mode frequencies, we make comparisons of our results with the recent experimental measurement [E. Haller \textit{et al.}, Science \textbf{325}, 1224 (2009)] and some previous theoretical predictions. We show that the experimental data are still not fully explained within current theoretical framework.
\begin{description}
\item[PACS numbers]
 67.85.-d, 02.60.Cb

\end{description}
\end{abstract}

\maketitle

\section{\label{sec:introduction}Introduction}
Many-particle systems in one dimension (1D) with a short-range interparticle interaction plays an important role in understanding fascinating quantum many-body physics~\cite{RevModPhys.83.1405,RevModPhys.85.1633}. For instance, a 1D interacting Bose gas is theoretically anticipated to experience different phases by changing interaction strength and temperature, and to exhibit a number of intriguing phenomena, such as effective fermionization and nontrivial quench dynamics~\cite{RevModPhys.83.1405,RevModPhys.85.1633}. Since the realization of Bose-Einstein condensation in 1995, more and more experiments have dealt with 1D atomic bosons in a harmonic trap at ultra-cold temperature, which can help test theoretical predictions and understand all these unusual phenomena. At present, many experiments have been conducted and controlled by means of Feshbach resonances~\cite{Haller04092009} or directly heating 1D quantum degenerate Bose gases~\cite{PhysRevLett.91.250402}, which measured and characterized various physical quantities, including momentum distribution~\cite{paredes2004tonks}, pair correlation~\cite{PhysRevLett.96.130403} and quenching rate~\cite{PhysRevLett.113.035301,cheneau2012light}.  Motivated by these rapid experimental advances, there are numerous theoretical studies based on sum-rule approach~\cite{PhysRevA.66.043610}, variational method~\cite{PhysRevA.90.013622}, local density approximation~\cite{2014arXiv1412.6855C} and diffusion Monte Carlo simulations~\cite{2014arXiv1412.4408G}. Experimental measured quantities, particularly momentum distribution~\cite{PhysRevA.85.031604} and pair correlation~\cite{PhysRevLett.91.040403,PhysRevA.71.053615,PhysRevLett.95.190406}, have been predicted and compared with experimental data.

In the recent experiment~\cite{Haller04092009}, the Feshbach resonance technique is used  to tune the interatomic interaction of a 1D Bose gas in a harmonic trap at extremely low temperature. The measured ratio of squared breathing mode frequency $\omega_{\rm{m}}^2/\omega_{\rm{ho}}^2$ exhibits a reentrant behaviour, from $4$ in the non-interacting regime to $3$ in the weakly interacting regime, and then back to $4$ in the Tonks-Girardeau regime~\cite{Girardeau1960}. Most recently, this interesting reentrant behaviour was addressed by two theoretical works based on simulations at zero temperature~\cite{2014arXiv1412.6855C,2014arXiv1412.4408G}. Choi and coworkers used a time-dependent modified nonlinear Schr\"{o}dinger equation (m-NLSE) with a local chemical potential replacing the conventional nonlinear term~\cite{2014arXiv1412.6855C}. Gudyma and collaborators combined a sum-rule approach in the mean-field regime and the local density approximation in the Tonks-Girardeau regime to describe the breathing mode~\cite{2014arXiv1412.4408G}. Diffusion Monte Carlo simulations for few particle numbers were performed, in order to obtain density profiles as inputs to the sum-rule approach. Comparing all results together, we find that there are still deviations between experimental data and theoretical predictions, particularly in the deep weakly-interacting regime. 

By considering finite particle number (i.e., varying in a range of $8\sim25$) and un-avoidable nonzero temperature in the realistic experiment~\cite{Haller04092009}, we therefore would like to address in a more systematic way their effects on collective mode frequencies of a 1D trapped Bose gas. We note that, at large number of particles, the temperature dependence of mode frequencies has been recently investigated by using a hydrodynamic theory~\cite{PhysRevA.90.013622}. 

In this paper, we investigate low-lying collective mode frequencies of a harmonically trapped Bose gas in 1D at a wide range of effective interaction parameter $\gamma_{\rm{_{eff}}}$, covering all interaction regimes from the non-interacting regime, mean-field regime to Tonks-Girardeau regime, by developing a generalized Bogoliubov theory at $T=0$ and a Hartree-Fock-Bogoliubov theory with Popov approximation at finite temperature~\cite{PhysRevB.53.9341,popov1991}. The former theory concentrates on the zero temperature case, where we follow the idea of the density-functional approach, use a generalized Gross-Pitaevskii equation and take the local chemical potential for uniform density obtained from the Lieb-Liniger model as the exchange energy (i.e., the nonliner term)~\cite{PhysRev.130.1605,*PhysRev.130.1616}. In the latter, the finite temperature effect is taken into account through the self-consistent Hartree-Fock-Bogoliubov equations in the weakly interacting regime. This is the only theory that we know so far to address both effects of nonzero temperature and finite particle number.

This paper is organized as follows. We provide details of the two theoretical methods in Sec.~\ref{sec:theoretical}. In Sec.~\ref{sec:result}, we present the ground state of 1D Bose gases (see Figs.~\ref{fig:1} and~\ref{fig:2}) and discuss the effects of particle number and temperature on excitation frequencies (see Figs.~\ref{fig:3},~\ref{fig:4} and~\ref{fig:5}). We then compare our results of the breathing mode frequency with  the experimental data~\cite{Haller04092009} and the previous theoretical predictions~\cite{2014arXiv1412.6855C} (see Figs.~\ref{fig:6} and~\ref{fig:7}). Mode frequencies of two higher collective modes are also discussed in Fig.~\ref{fig:8}. Summary and outlook are given in Sec.~\ref{sec:summary}. 

\section{Theoretical Framework\label{sec:theoretical}}

\subsection{A density functional method with Lieb-Liniger integrals at zero temperature\label{subsectionA}}
We start with a 1D atomic Bose gas with a repulsive zero-range potential. The system of $N$ particles can be described  with the Lieb-Liniger Hamiltonian~\cite{PhysRev.130.1605,*PhysRev.130.1616},
\begin{equation}
\mathcal{H}=-\frac{\hbar^{2}}{2m}\sum\limits_{i=1}^N\frac{\partial^{2}}{\partial x_{i}^{2}}+g_{\rm{_{1D}}}\sum\limits_{i<j}^N\delta(x_{i}-x_{j}),\label{eq:hamiltonian1}
\end{equation}
where $m$ is the mass of atom and $g_{\rm{_{1D}}}$ is the 1D interaction strength of form 
\begin{equation}
g_{\rm{_{1D}}}=\frac{2\hbar^2}{ma_{\rm{_{1D}}}},
\end{equation} 
characterizing the interaction between bosons. $a_{\rm{_{1D}}}$ is the 1D scattering length calculated by~\cite{PhysRevLett.81.938}
\begin{equation}
a_{\rm{_{1D}}}=-\frac{a_{\perp}^2}{2a_{\rm{_{3D}}}}\left[ 1-\mathcal{C} \frac{a_{\rm{_{3D}}}}{a_{\perp}} \right], \label{eq:CIR}
\end{equation}
where $a_{\rm{_{3D}}}$ is the three-dimensional (3D) scattering length, $a_{\perp}=[\hbar/(m\omega_{\perp})]^{1/2}$ is a two-dimensional (2D) harmonic oscillator characteristic length (see below) and the constant $\mathcal{C}\simeq1.4603$. Experimentally, systems of 1D Bose gases are usually trapped in a cylindrically symmetric potential with strong transverse confinement and weak longitudinal confinement (i.e., with trapping frequencies $\omega_{\perp}\gg\omega_x$). Transverse excitations are not taken into account if the transverse vibrational energy $\hbar\omega_{\perp}$ is much larger than the chemical potential or the energy scale  of the thermal cloud ($\hbar\omega_{\perp}\gg\mu,k_{\rm{B}}T$)~\cite{Haller04092009, PhysRevLett.91.250402, paredes2004tonks, PhysRevLett.96.130403, PhysRevLett.113.035301}. However, the scattering of two atoms in the lowest transverse mode could strongly be affected by high transverse excitations, when the 3D scattering length $a_{\rm{_{3D}}}$ is close to the 2D harmonic oscillator length $a_{\perp}$.  This leads to a confinement-induced Feshbach resonance, as can be seen in Eq.~\eqref{eq:CIR}.

\subsubsection{\label{subsection_LL}Lieb-Liniger model}
Lieb and Liniger investigated this system at $T=0$ and solved it exactly in 1960s. They dealt with the Hamiltonian~\eqref{eq:hamiltonian1} and the associated Schr\"{o}dinger equation by means of Bethe ansatz~\cite{Betha1931}, and obtained exactly the ground state as well as low-lying excited states in a uniform gas with a constant density $n=N/V$~\cite{PhysRev.130.1605,*PhysRev.130.1616}.

In the Lieb-Liniger model, they defined a dimensionless interaction parameter $\gamma(n)$ as a function of $g_{\rm{_{1D}}}$, which is
\begin{equation}
\gamma(n)\equiv\frac{mg_{\rm{_{1D}}}}{\hbar^2n}=\frac{2}{n a_{\rm{_{1D}}}}.
\end{equation}
By solving exactly the Hamiltonian, they got a group of integral equations (i.e., Lieb-Liniger integrals), including the normalization condition and the equation of the ground state energy. The ground state energy has the form
\begin{equation}
E_0=\frac{N\hbar^2n^2}{2m}e(\gamma),
\end{equation}
where $e(\gamma)$ is a dimensionless function of $\gamma$, which can be obtained by solving the following Lieb-Liniger integrals,
\begin{equation}
\begin{aligned}
g(x)&=\frac{1}{2\pi}+\frac{1}{2\pi}\int_{-1}^{1}\frac{2\lambda}{(x-x^{\prime})^{2}+\lambda^{2}}g(x^{\prime})dx^{\prime},\\
\lambda&=\gamma\int_{-1}^{1}g(x)dx,\\
e(\gamma)&=(\frac{\gamma}{\lambda})^{3}\int_{-1}^{1}x^{2}g(x)dx\label{eq:LL}.
\end{aligned}
\end{equation}
In the integrals, $g(x)$ is the distribution function of the quasi-momentum, which is represented by the variable $x$ after a rescaling~\cite{PhysRev.130.1605}. Thus, the variables $x,x^{\prime}$ are all bounded in the range $[-1,1]$, as shown in the upper and lower limits of the integrals. Variable $\lambda$ is proportional to $\gamma$ since the remaining integral part in $\lambda$ is fixed. Hence, $\lambda$ is also proportional to the interaction strength $g_{\rm{_{1D}}}$ for a fixed uniform density $n$.

By numerically solving the integrals, it can be shown that $e(\gamma)$ is a monotonically increasing function of $\gamma$~\cite{PhysRev.130.1605}. In the limit of $\gamma=0$, $e(0)=0$ and the ground state energy is $E_0=0$, corresponding to the case that all free bosons occupy the zero-momentum state. When $\gamma$ is sufficiently large, the asymptotic value of $e(\gamma)$ is $\pi^2/3$, which exactly coincides the predicted value for impenetrable bosons in 1D by Girardeau~\cite{Girardeau1960}. All the information of the ground state of 1D Bose gas can then be calculated from $e(\gamma)$. The energy per particle at the ground state is $\epsilon(n)=E_0/N=\hbar^2n^2e(\gamma)/(2m)$, and the corresponding chemical potential $\mu(n)$ is given by
\begin{equation}
\mu(n)=\frac{\partial(n\epsilon(n))}{\partial n}=\frac{\hbar^{2}n^{2}}{2m}\mu(\gamma).
\end{equation}
It is straightforward to see that the dimensionless chemical potential $\mu(\gamma)$ can be calculated by 
\begin{equation}
\mu(\gamma)=3e(\gamma)-\gamma e^{\prime}(\gamma). 
\end{equation}
In Ref.~\cite{2014arXiv1412.6855C}, Choi and coworkers proposed the following analytic expression for the chemical potential at large $\gamma$,
\begin{equation}
\mu(n)\approx\frac{\pi^2}{2}\frac{\hbar^2 n^2}{m} \left[ \frac{\gamma^2(n)(2+3\gamma(n))}{3(2+\gamma(n))^3} \right],
\end{equation}
near the Tonks-Girardeau limit. Compared with the numerical results of Lieb-Liniger integrals, this expression turns out to be very accurate, as long as $\gamma$ is larger than $10$.

\subsubsection{Generalized Gross-Pitaevskii theory}
The Gross-Pitaevskii theory provides a good description of weakly-interacting atomic Bose-Einstein condensates at $T=0$. However, it is known that the theory fails in low dimensions ($d\leq2$) and therefore needs appropriate modifications. In Ref.~\cite{PhysRevLett.85.1146}, Kolomeisky and coworkers suggested an interesting modification in the Tonks-Girardeau limit for 1D trapped Bose gases. Here, we generalize their idea to all interaction strengths, following the procedure of using an improved exchange energy in the density-function approach.

Recall that the standard Gross-Pitaevskii equation with a condensate order parameter (wavefunction) $\Psi(x,t)$ is given by
\begin{equation}
i\hbar\frac{\partial\Psi(x,t)}{\partial t}=\left[ \mathcal{H}_{0}+g_{\rm{_{1D}}}n(x,t) \right]\Psi(x,t)\label{eq:GP1}
\end{equation}
with the single particle Hamiltonian 
\begin{equation}
\mathcal{H}_{0}=-\frac{\hbar^{2}}{2m}\frac{\partial^{2}}{\partial x^{2}}+V_{\rm{ext}}(x),
\end{equation}
where $V_{\rm{ext}}(x)=\frac{1}{2}m\omega_{\rm{ho}}^{2}x^{2}$ is the harmonic trapping potential and $n(x,t)=|\Psi(x,t)|^{2}$ is the particle density. The Hartree term $g_{\rm{_{1D}}}n(x,t)$ in Eq.~\eqref{eq:GP1} is only applicable in the weak coupling regime.

Away from the weak coupling regime ($na_{\rm{_{1D}}}\gg1$ or $\gamma\ll1$), we may use local density approximation (LDA) to determine the ground state of a trapped system. The LDA amounts to setting,
\begin{equation}
\mu=\mu_{\rm{loc}}(n(x))+V_{\rm{ext}}(x), \label{eq:lda}
\end{equation}
with $\mu$ being the global chemical potential. Once we know the local chemical $\mu_{\rm{loc}}(n)$ for a uniform Bose gas from the Lieb-Liniger model, we can then determined the density profile $n(x)$ by inversely solving Eq.~\eqref{eq:lda}.The important observation made by Kolomeisky and coworkers is that in the Tonks-Girardeau limit, one may simply obtain a modified Gross-Pitaevskii equation by using $\mu_{\rm{loc}}(n)=\pi^2\hbar^2n^2/2m$ to replace the Hartree term $g_{\rm{_{1D}}}n$~\cite{PhysRevLett.85.1146}. Motivated by this work, one can use $\mu_{\rm{loc}}(n)$ calculated numerically in the Lieb-Liniger model, to obtain the corresponding result in the intermediate regime between the mean-field limit and the Tonks-Girardeau limit.

Modifying the interaction term $g_{\rm{_{1D}}}n$ to the local chemical potential $\mu_{\rm{loc}}(n)$, the generalized Gross-Pitaevskii equation reads,
\begin{equation}
i\hbar\frac{\partial\Psi(x,t)}{\partial t}=\left[ \mathcal{H}_{0} +\mu_{\rm{loc}}\left( n(x,t) \right) \right] \Psi(x,t), \label{eq:GP3}
\end{equation}
which has the same form as the m-NLSE equation adopted by Choi and coworkers~\cite{2014arXiv1412.6855C}. 

The idea of directly modifying the exchange-energy-like term at zero temperature is supported by the following derivation of hydrodynamic equation. That is, we rewrite the order parameter as 
\begin{equation}
\Psi(x,t)=n^{\frac{1}{2}}(x,t)e^{i\mathbf{\theta}(x,t)},
\end{equation}
where $n(x,t)$ is now interpreted as the time-dependent \emph{superfluid density} of the system and $\theta(x,t)$ is the associated phase. We therefore introduce a superfluid velocity field $\mathbf{v}(x,t)=\frac{\hbar}{m}\frac{\partial \mathbf{\theta}(x,t)}{\partial x}\mathbf{e}_x$ and rewrite the generalized Gross-Pitaevskii equation in terms of the superfluid density $n(x,t)$ and superfluid velocity $\mathbf{v}(x,t)$:
\begin{equation}
\begin{aligned}
m\frac{\partial \mathbf{v}}{\partial t}+\frac{\partial}{\partial x}(\mu_{\rm{loc}}(n)+V_{\rm{ext}}(x)+\frac{1}{2}m\mathbf{v}^2)&=0,\\
\frac{\partial n}{\partial t}+\frac{\partial}{\partial x}(n\mathbf{v})&=0.\\
\end{aligned}
\end{equation}
Here we have neglected a quantum pressure term, which is small in the long wave-length limit. Thus, it is clear that our generalized Gross-Pitaevskii equation is identical to the standard 1D time-dependent hydrodynamic equations in the long wave-length limit. For a finite particle number system, it is preferable to use the generalized Gross-Pitaevskii equation to understand the dynamics of interacting 1D Bose gases. The effect of finite particle number is taken into account by the quantum pressure term that is neglected in the hydrodynamic equations.

\subsubsection{Generalized Bogoliubov theory}
We now consider excited states of 1D Bose gases, which can be treated as small oscillations around the superfluid density at the ground state (i.e., $\phi_0(x)$) with frequencies $\omega_i$. Their wavefunctions, $u_i$ and $v_i$, are given by~\cite{RevModPhys.71.463}, 
\begin{equation}
\Psi(x,t)=e^{-i\mu t/\hbar} \left[ \phi_0(x)+u_i(x)e^{-i\omega_i t/\hbar}+v^*_i(x)e^{i\omega_i t/\hbar} \right].\label{eq:smalloscillation}
\end{equation}
The corresponding density $n=|\Psi(x,t)|^2$ is
\begin{equation}
n=|\phi_0|^2+\left[\phi^*_0(u_ie^{-i\omega_i t/\hbar}+v^*_ie^{i\omega_i t/\hbar})+ \text{H.c.}\right],
\end{equation}
where we keep only the linear terms in the complex functions $u_i$ and $v_i$. Accordingly, the local chemical potential $\mu_{\rm{loc}}(n)$ can be written as
\begin{equation}
\mu_{\rm{loc}}(n)=\mu_{\rm{loc}}(n_0)+\frac{\partial\mu_{\rm{loc}}}{\partial n} \left[ e^{-i\omega_i t/\hbar}(\phi^*_0 u_i+\phi_0 v_i)+\text{H.c.} \right]\label{eq:localmu}
\end{equation}
in the Taylor expansion with $n_0(x)=|\phi_0(x)|^2$, where higher orders are neglected. Taking Eqs.~\eqref{eq:smalloscillation} and~\eqref{eq:localmu} back into Eq.~\eqref{eq:GP3} and sorting out the terms in $e^{-i\mu t/\hbar}$, $e^{-i(\mu+\hbar\omega_i) t/\hbar}$, $e^{-i(\mu-\hbar\omega_i) t/\hbar}$, one obtains respectively the static generalized Gross-Pitaevskii equation,
\begin{equation}
\mu\phi_{0}(x)=\left[\mathcal{H}_{0}+\mu_{\rm{loc}}(n_{0}) \right]\phi_{0}(x),\label{eq:static1}
\end{equation}
as well as the coupled Bogoliubov equations
\begin{equation}
\begin{aligned}
\mathcal{L} u_i(x)+ \mathcal{M} v_i(x)=&+\hbar\omega_iu_i(x),\\
\mathcal{L} v_i(x)+ \mathcal{M} u_i(x)=&-\hbar\omega_iv_i(x),\label{eq:coupled}
\end{aligned}
\end{equation}
where we have defined the operators
\begin{equation}\mathcal{L}=\mathcal{H}_{0}-\mu+\mu_{\rm{loc}}(n_{0})+\left[\frac{\partial\mu_{\rm{loc}}(n)}{\partial n} \right]_{n=n_0}n_{0},
\end{equation}
and 
\begin{equation}
\mathcal{M}=\left[\frac{\partial\mu_{\rm{loc}}(n)}{\partial n} \right]_{n=n_0} n_{0}.
\end{equation}
The Bogoliubov wave-functions $u_i(x)$ and $v_i(x)$ satisfy the normalization condition,
\begin{equation}
\int_{-\infty}^{\infty}dx(u_i^{*}(x)u_j(x)-v_i^{*}(x)v_j(x))=\delta_{ij}.
\end{equation}

The above formalism (with $\mu_{\rm{loc}}(n)=g_{\rm{_{1D}}}n$) was first introduced by Pitaevskii, in order to investigate excitations of a vortex line in a uniform Bose gas. One can get the same result if one diagonalizes the Hamiltonian with Bogoliubov transformation~\cite{RevModPhys.71.463}. In 1996, Burnett and colleagues used the coupled equations~\eqref{eq:coupled} to study the properties of excited states in 3D weakly-interacting trapped Bose gases~\cite{edwards1996zero}.

After numerically solving the static Gross-Pitaevskii equation~\eqref{eq:static1} and the coupled equations~\eqref{eq:coupled}, one can obtain directly the profile of the ground state, as well as excited states with  energies $\epsilon_i=\hbar\omega_i$ and hence the frequency $\omega_\text{m}$ of the breathing or compressional monopole mode.

\subsection{Hartree-Fock-Bogoliubov method with Popov approximation at finite temperature\label{subsectionB}}
In quantum many-body systems, the Hartree-Fock approximation (or self-consistent approximation) provides a good description of the ground state wavefunction and energy at weak couplings. It works for both bosons and fermions. It can also be generalized to finite temperature to solve thermodynamics~\cite{RevModPhys.71.463}. For a 3D weakly interacting Bose gas, Shi and Griffin have shown that the Hartree-Fock theory is useful at finite temperature except a small critical area near the transition temperature~\cite{Shi1997,Shi19981}.

In the case of a 1D trapped Bose gas, the grand-canonical Hamiltonian takes the form~\cite{fetter2003quantum,feenberg2012theory}
\begin{widetext}
\begin{equation}
\hat{\mathcal{H}}^{\rm{GC}}=\int dx\hat{\Psi}^{\dagger}(x,t)(\mathcal{H}_{0}-\mu)\hat{\Psi}(x,t)
+\frac{1}{2}\iint dxdx^{\prime}\hat{\Psi}^{\dagger}(x,t)\hat{\Psi}^{\dagger}(x^{\prime},t)U(x-x^{\prime})\hat{\Psi}(x^{\prime},t)\hat{\Psi}(x,t),
\end{equation}
\end{widetext}
where $\hat{\Psi}(x,t)$ is the bosonic field operator. By taking a $\delta$-function interaction $U(x-x^{\prime})=g_{\rm{_{1D}}}\delta(x-x^{\prime})$ between bosons, the Heisenberg equation of motion is given by
\begin{equation}
\begin{aligned}
i\hbar\frac{\partial\hat{\Psi}(x,t)}{\partial t}&=\left[\hat{\Psi}(x,t),\hat{\mathcal{H}}^{\rm{GC}}\right]\\
       &=(\mathcal{H}_0-\mu)\hat{\Psi}(x,t)+g_{\rm{_{1D}}}\hat{\Psi}^{\dagger}\hat{\Psi}\hat{\Psi}(x,t).\label{eq:Heisenberg}
\end{aligned}
\end{equation}
We separate the operator $\hat{\Psi}(x,t)$ into the condensate part and non-condensate part~\cite{Fetter197267,edwards1996zero,PhysRevLett.77.1671,PhysRevA.57.R32}: 

\begin{equation}
\hat{\Psi}(x,t)=\Phi(x,t)+\hat{\eta}(x,t),
\end{equation}
which leads to $\langle\hat{\Psi}\rangle=\Phi$ and $\langle\hat{\eta}\rangle=0$. Thus, we obtain $\langle\hat{\Psi}^{\dagger}\hat{\Psi}\rangle=|\Phi|^2+\langle\hat{\eta}^{\dagger}\hat{\eta}\rangle$ and $\langle\hat{\Psi}\hat{\Psi}\rangle=\Phi^2+\langle\hat{\eta}\hat{\eta}\rangle$.

Within the mean-field approximation, we rewrite the term $\hat{\Psi}^{\dagger}\hat{\Psi}\hat{\Psi}$ as~\cite{PhysRevB.53.9341}
\begin{equation}
\langle\hat{\Psi}^{\dagger}\hat{\Psi}\hat{\Psi}\rangle
=(|\Phi|^2+2\langle\hat{\eta}^{\dagger}\hat{\eta}\rangle)\Phi
+\langle\hat{\eta}\hat{\eta}\rangle\Phi^*.\label{eq:mfapproxi}
\end{equation}
By taking an average on both sides of Eq.~\eqref{eq:Heisenberg} and making use of Eq.~\eqref{eq:mfapproxi}, we then find that
\begin{equation}
i\hbar\frac{\partial\Phi}{\partial t}=(\mathcal{H}_0-\mu)\Phi+g_{\rm{_{1D}}}\left[ (n_c+2n_t)\Phi+n_a\Phi^* \right],\label{eq:average}
\end{equation}
where $n_c\equiv|\Phi(x,t)|^2$ is the time-dependent condensate density, $n_t\equiv\langle\hat{\eta}^{\dagger}\hat{\eta}\rangle$ is the non-condensate thermal density and $n_a\equiv\langle\hat{\eta}\hat{\eta}\rangle$ is the anomalous thermal density.

The equation of motion for the non-condensate operator $\hat{\eta}$ can be similarly obtained:
\begin{equation}
i\hbar\frac{\partial\hat{\eta}}{\partial t}=(\mathcal{H}_0-\mu)\hat{\eta}+g_{\rm{_{1D}}} \left[2(n_c+n_t)\hat{\eta}+(n_c+n_a)\hat{\eta}^{\dagger} \right].
\end{equation}
To solve this equation, we then use the Bogoliubov transformation to expand the non-condensate operator $\hat{\eta}(x,t)$ and its conjugate as
\begin{equation}
\begin{aligned}
& \hat{\eta}(x,t)  =e^{-i\tilde{\mu} t/\hbar}\underset{i}{\sum} \left[u_{i}(x)e^{-i\omega_{i}t}\hat{\alpha}_{i}+v_{i}^{*}(x)e^{i\omega_{i}t}\hat{\alpha}_{i}^{\dagger}\right],\\
& \hat{\eta}^{\dagger}(x,t)  =e^{i\tilde{\mu} t/\hbar}\underset{i}{\sum} \left[v_{i}(x)e^{-i\omega_{i}t}\hat{\alpha}_{i}+u_{i}^{*}(x)e^{i\omega_{i}t}\hat{\alpha}_{i}^{\dagger}\right],\label{eq:bogoliubov2}
\end{aligned}
\end{equation}
i.e., the non-condensate operator $\hat{\eta}(x,t)$ is rewritten in a quasi-particle basis. The operators $\hat{\alpha}^{\dagger}$ and $\hat{\alpha}$ are the creation and annihilation operators for quasi-particles respectively and, $u_i$ and $v_i$ are the corresponding amplitudes. Bogoliubov quasi-particles are assumed to be non-interacting, and their operators $\hat{\alpha}^{\dagger}$ and $\hat{\alpha}$ satisfy the bosonic commutation relations: 
\begin{equation}
\left[\hat{\alpha}^{\dagger}_i,\hat{\alpha}_j \right]=\delta_{ij},~\left[\hat{\alpha}^{\dagger}_i,\hat{\alpha}^{\dagger}_j\right]=\left[\hat{\alpha}_i,\hat{\alpha}_j\right]=0.
\end{equation}

Taking the Popov approximation, neglecting the anomalous term with $n_a\equiv\langle \hat{\eta}\hat{\eta}\rangle$~\cite{popov1991}, we finally obtain the time-dependent Hartree-Fock-Bogoliubov equations (HFBP)~\cite{PhysRevB.53.9341,PhysRevA.57.R32,PhysRevLett.95.180401}, which consists of two parts: 

(A) a modified Gross-Pitaevskii equation 
\begin{equation}
i\hbar\frac{\partial\Phi}{\partial t}=\left[\mathcal{H}_0-\mu+g_{\rm{_{1D}}}(n_c+2n_t)\right]\Phi,\label{eq:GP2}
\end{equation}
which leads to
\begin{equation}
\tilde{\mu}\phi(x)=\left[\mathcal{H}_0+g_{\rm{_{1D}}}(n_{0}(x)+2n_t(x))\right]\phi(x,t).\label{eq:finalstatic}
\end{equation}
Here $\Phi(x,t)=e^{-i\Delta\mu t/\hbar}\phi(x)$ and $\tilde{\mu}=\Delta\mu+\mu$. We have introduced a small chemical potential difference $\Delta\mu$ to allow a finite particle number of condensate (see Eq.~\eqref{eq:N0} below). The static condensate density is $n_{0}\equiv|\phi(x)|^2$ and the thermal density is $n_t(x)\equiv\langle\hat{\eta}^{\dagger}\hat{\eta}\rangle$.

(B) coupled Bogoliubov equations
\begin{equation}
\left[\begin{array}{cc}
\mathcal{L} & \mathcal{M}\\
\mathcal{M} & \mathcal{L}
\end{array}\right]\left[\begin{array}{cc}
u_{i}(x)\\
v_{i}(x)
\end{array}\right]=\hbar\omega_{i}\left[\begin{array}{cc}
+u_{i}(x)\\
-v_{i}(x)
\end{array}\right],\label{eq:bdg2}
\end{equation}
which are obtained by substituting Eq.~\eqref{eq:bogoliubov2} back into the equation of motion for the non-condensate operator $\hat{\eta}$. 
In the coupled equations, we have defined 
\begin{equation}
\mathcal{L}=\mathcal{H}_{0}-\tilde{\mu}+2g_{\rm{_{1D}}}\left( n_{0}(x)+n_t(x) \right),
\end{equation}
and 
\begin{equation}
\mathcal{M}=g_{\rm{_{1D}}} n_{0}(x).
\end{equation}

In the modified Gross-Pitaevskii equation, there are condensate and non-condensate densities in the exchange term, both of which have to be solved self-consistently. It should be noted that, for the condensate density, because our bosonic system has a finite particle number, we have to introduce a small chemical potential difference to account for the finite condensate particle number $N_0=\int dxn_0(x)$, i.e.,
\begin{equation}
N_{0}= \frac{1}{e^{\beta\Delta\mu}-1},\label{eq:N0}
\end{equation}
with the inverse temperature $\beta=1/k_{\rm{B}}T$. The quasi-particle occupation number is affected by the finite condense number as well. For the $i$-th quasi-particle occupation number $N_{i}\equiv\langle\hat{\alpha}^{\dagger}_i\hat{\alpha}_i\rangle$, we have, 
\begin{equation}
\begin{aligned}
N_{i}&=\frac{1}{e^{\beta(\hbar\omega_{i}+\Delta\mu)}-1}\\
     &=\frac{1}{(1+\frac{1}{N_{0}})e^{\beta\hbar\omega_{i}}-1}.
\end{aligned}
\end{equation}
Therefore, the thermal density is given by
\begin{equation}
\begin{aligned}
n_t(x)&\equiv \langle\hat{\eta}^{\dagger}\hat{\eta}\rangle\\
        &=\sum\limits_{i}\left[u_{i}^2(x)+v_{i}^2(x)\right]\frac{1}{(1+\frac{1}{N_{0}})e^{\beta\hbar\omega_{i}}-1}+\sum\limits_{i}v_{i}^2(x)\\
        &=\sum\limits_{i}\left[u_{i}^2(x)+v_{i}^2(x)\right]N_{i}+\sum\limits_{i}v_{i}^2(x).
\end{aligned}
\end{equation}
The chemical potential of the system, $\mu$, is to be determined by the number equation for the total number of atoms, 
\begin{equation}
N=N_0+N_T, 
\end{equation}
where $N_T=\int dxn_t(x)$.

\section{Results and Discussions\label{sec:result}}
We are now ready to perform numerical calculations of the above mentioned approaches and compare our results with the experiment data~\cite{Haller04092009} and the previous theoretical predictions~\cite{2014arXiv1412.6855C}. In the experiment, a 2D optical lattice is used to trap about $(1\sim4)\times10^4$ Cs atoms in $(3\sim6)\times10^3$ 1D tubes with $8\sim25$ atoms in the center tube. Choi \textit{et al.} dealt with the case of a particle number $N=25$ and introduced an effective dimensionless interaction parameter $\gamma_{\rm{_{eff}}}$, which is defined as
\begin{equation}
\gamma_{\rm{_{eff}}}\equiv\frac{2}{n_{\rm{_{TG}}}(0)| a_{\rm{_{1D}}}|}=\frac{g_{\rm{_{1D}}}\pi}{\sqrt{2N}}. \label{eq:gamma_eff}
\end{equation}
Here, $n_{\rm{_{TG}}}(0)=\sqrt{2Nm\omega_{\rm{ho}}/\hbar}/\pi$ is the peak density in the Tonks-Girardeau regime at the tube center~\cite{PhysRevLett.86.5413}.

In our calculations, harmonic oscillator units are used with $\hbar=\omega_{\rm{ho}}=m=1$ and $k_{\rm{B}}=1$. Length and energy are written in the units of harmonic oscillator length $a_{\rm{ho}}=[\hbar/(m\omega_{\rm{ho}})]^{1/2}$ and harmonic oscillator energy $\hbar\omega_{\rm{ho}}=[\hbar^2/(ma_{\rm{ho}}^2)]$, respectively. 

We fix the particle number at $N=8, 17, 25$, and vary the effective interaction parameter $\gamma_{\rm{_{eff}}}$. The ground state and collective modes of the 1D harmonically trapped Bose gas at $T=0$ are obtained, by numerically solving the generalized Gross-Pitaevskii equation~\eqref{eq:static1}, and the generalized Bogoliubov equations (\ref{eq:coupled}), where the local chemical potential is obtained with Lieb-Liniger integrals~\cite{PhysRev.130.1605,*PhysRev.130.1616}. Then we introduce the non-condensate thermal density $n_t$ by means of the HFBP theory, and compare the ground states and collective behaviours at finite $T$ with the result at $T=0$.

\subsection{Density profile}
In this subsection, we study the density profile of the ground states. In particular, at zero temperature we compare the results obtained by the generalized Gross-Pitaevskii theory with those predicted by the HFBP theory.

\begin{figure}[t]
\centering
\includegraphics[scale=0.35]{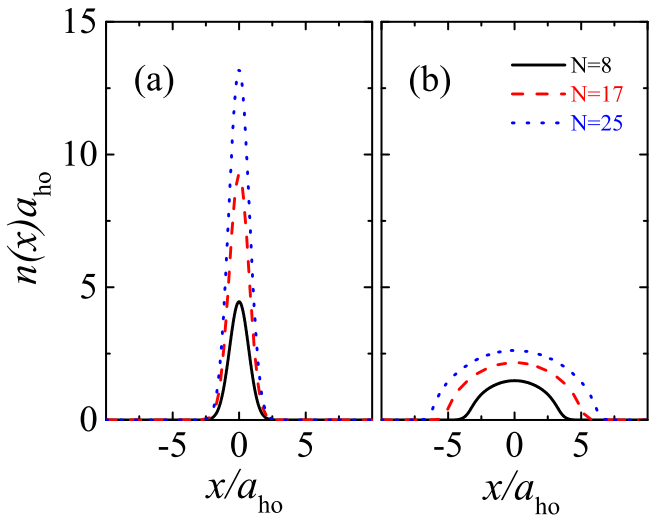}
\caption{\label{fig:1} (Color online) Density profiles in the mean-field regime (a) and in the Tonks-Girardeau regime (b), where the effective interaction parameter $\gamma_{\rm{_{eff}}}=10^{-2}$ and $10$ have been used, respectively. The results are calculated with $N=8$ (black solid lines), $N=17$ (red dashed lines) and $N=25$ (blue dotted lines). The density $n(x)$ and position $x$ are taken in units of harmonic oscillator length $a_{\rm{_{ho}}}=\sqrt{\hbar/(m\omega_{\rm{ho}})}$ and $a_{\rm{ho}}^{-1}$, respectively.}
\end{figure}

\subsubsection{$T=0$ case: the generalized Gross-Pitaevskii theory}

The ground states for different particle number $N=8,17,25$ at $T=0$ are shown in Fig.~\ref{fig:1}, obtained by solving the generalized Gross-Pitaevskii theory. 
In the figure, we focus on the mean-field regime (with $\gamma_{\rm{_{eff}}}=10^{-2}$, the left panel) and the Tonks-Girardeau regime (with $\gamma_{\rm{_{eff}}}=10$, the right panel). At $\gamma_{\rm{_{eff}}}=10^{-2}$, the density profile is roughly a Gaussian curve, while at $\gamma_{\rm{_{eff}}}=10$ it tends to be a semicircle. In both regimes, for a given $\gamma_{\rm{_{eff}}}$, the height of density profiles is enhanced with increasing particle number.

\begin{figure}[t]
\centering
\includegraphics[scale=0.35]{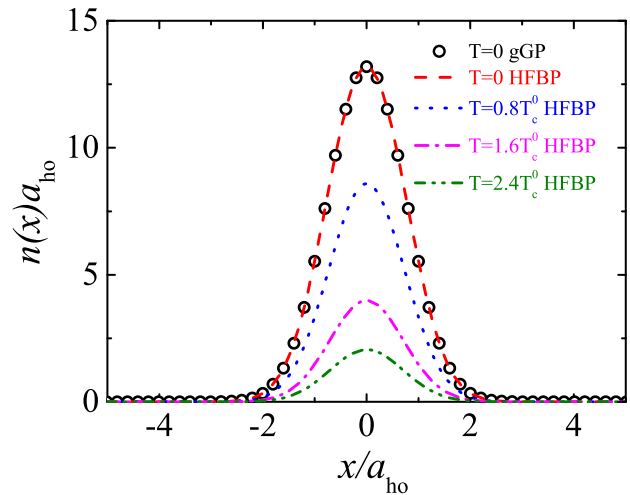}
\caption{\label{fig:2} (Color online) Density profile: at $T=0$ predicted by the generalized Gross-Pitaevskii theory (black circles); at $T=0$ (red dashed lines) and finite temperatures (other colorful lines) calculated by the HFBP theory. The results are shown at the effective interaction parameter $\gamma_{\rm{_{eff}}}=10^{-2}$ in the mean-field regime. The particle number $N$ is fixed at $25$. $T_{\rm{c}}^0$ is the critical temperature for a 1D ideal Bose gas, which can be estimated as $k_{\rm{B}}T_{\rm{c}}^0=\hbar\omega_{\rm{ho}}N/\ln(2N)$~\cite{PhysRevA.54.656,RevModPhys.71.463}.}
\end{figure}

\subsubsection{Finite $T$ at weak couplings: the HFBP theory\label{sec:finiteT}}

The 1D HFBP theory is valid only in the weak coupling regime. The corresponding density profiles at different temperatures are shown in Fig.~\ref{fig:2}. For $T=0$, the density profile predicted by the generalized Gross-Pitaevskii theory (indicated as gGP in the figure) is also presented by black circle for comparison. It agrees well with the prediction of the HFBP theory (i.e., the red dashed line). With increasing temperature, the condensate fraction decreases, as well as the condensate occupation number, leading to the decreasing of the height of density profiles.

\subsection{Breathing mode}
In this subsection, collective modes, especially the breathing mode, are investigated with the generalized Bogoliubov theory and the HFBP theory. The dipole mode frequency should precisely be the trapping frequency $\omega_{\text{ho}}$, according to the Kohn theorem. We recover the result with a relative error about $0.1\%\sim0.2$\% with respect to $\omega_{\text{ho}}$.
 
\subsubsection{Particle number effect in two limits\label{sec:particleN}}
The experiment on the breathing mode frequency was conducted for particle numbers in the range of $8\sim25$. There is a deviation between the experimental results~\cite{Haller04092009}  and one of previous numerical simulations at $N=25$ in the weak coupling regime~\cite{2014arXiv1412.6855C}. In order to check whether this is due to the effect of different particle number, we calculate the breathing mode frequency with different $N$.

\begin{figure}[t]
\centering
\includegraphics[scale=0.30]{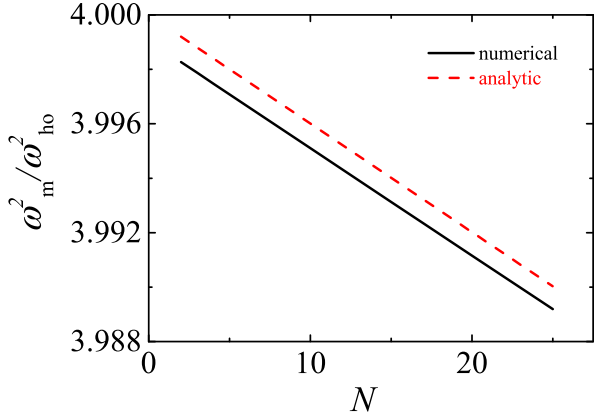}
\caption{\label{fig:3} (Color online) The ratio of the squared breathing mode frequency $\omega_{\rm{m}}^2/\omega_{\rm{ho}}^2$ as a function of the particle number $N$. Our numerical calculation is shown by the black solid line, and the analytic result Eq.~\eqref{eq:neffect_gp} is shown by the red dashed line. All results are near the non-interacting limit with an interaction strength $g_{\rm{_{1D}}}=10^{-3}$ in the trap units.} 
\end{figure}

In the deep weak coupling limit, it is convenient to define a Hartree parameter $\lambda=|a_{\rm{_{1D}}}|/(Na_{\rm{ho}})\gg1$. The sum-rule approach predicts that in the limit of $\lambda\gg1$, the correction of a finite particle number on the squared breathing mode frequency is~\cite{2014arXiv1412.4408G}
\begin{equation}
\begin{aligned}
\frac{\omega_{\rm{m}}^2}{\omega_{\rm{ho}}^2}&\simeq4(1-c_{\rm{_N}}\lambda^{-1})\\
                                  &=4-\frac{g_{\rm{_{1D}}}}{\sqrt{2\pi}} N, \lambda\rightarrow\infty\label{eq:neffect_gp}
\end{aligned}
\end{equation}
where $c_{\rm{_N}}=1/\sqrt{8\pi}$ for all $N\geq2$. This analytic prediction indicates that the squared breathing mode frequency ratio $\omega_{\rm{m}}^2/\omega_{\rm{ho}}^2$ is proportional to the particle number $N$ at a fixed $g_{\rm{_{1D}}}$.

In Fig.~\ref{fig:3}, we show our numerical results (solid line) at $\lambda=|a_{\rm{_{1D}}}|/(Na_{\rm{ho}})\gg1$, with the effective interaction parameter $\gamma_{\rm{_{eff}}}$ varying from $4.4\times10^{-4}$ to $1.6\times10^{-3}$ at a constant $g_{\rm{_{1D}}}=10^{-3}$. The analytic results Eq.~\eqref{eq:neffect_gp} are also shown by the dashed line. There is a very good agreement, within a relative error $0.1\%$. Presumably, this small discrepancy is due to the over-estimation of the mode frequency in the sum-rule approach, which predicts only an upper bound for the mode frequency~\cite{PhysRevA.68.043610}. 

In the opposite limit of strong couplings, which is characterized by the parameter $\Lambda=Na_{\rm{_{1D}}}^2/a_{\rm{ho}}^2\ll1$, the correction on the breathing mode frequency due to a finite $N$ is also known:~\cite{PhysRevA.89.063616}
\begin{equation}
\begin{aligned}
\frac{\omega_{\rm{m}}^2}{\omega_{\rm{ho}}^2}&\simeq4(1-C_{\rm{_N}}\sqrt{\Lambda})\\
                                  &=4-\frac{8}{g_{\rm{_{1D}}}} C_{\rm{_N}}\sqrt{N}, \Lambda\rightarrow0\label{eq:neffect_tg}
\end{aligned}
\end{equation}
where  $C_{\rm{_N}}$ is given for all $N\geq2$ by
\begin{equation}
\begin{aligned}
C_{\rm{_N}}=&\frac{3\sqrt{2N}}{\pi\sqrt{\pi}}\frac{\Gamma(N-\frac{5}{2})\Gamma(N+\frac{1}{2})}{\Gamma(N)\Gamma(N+2)}{}\\
      &\times_3F_2\left(\frac{3}{2},1-N,-N;\frac{7}{2}-N,\frac{1}{2}-N;1\right).\label{eq:CN}
\end{aligned}
\end{equation}
The $N$-dependence of $C_{\rm{_N}}$ is very weak and its value varies from $C_{_2}\approx0.282$ to $C_{_{\infty}}\approx0.306$, and to $C_{_{25}}\approx0.305$ in our case. It is clear from Eq.~\eqref{eq:neffect_tg} that for a fixed $g_{\rm{_{1D}}}$, the ratio of the squared breathing mode frequency $\omega_{\rm{m}}^2/\omega_{\rm{ho}}^2$ has a linear dependence on the combined variable $C_{\rm{_N}}\sqrt{N}$.

\begin{figure}[t]
\centering
\includegraphics[scale=0.30]{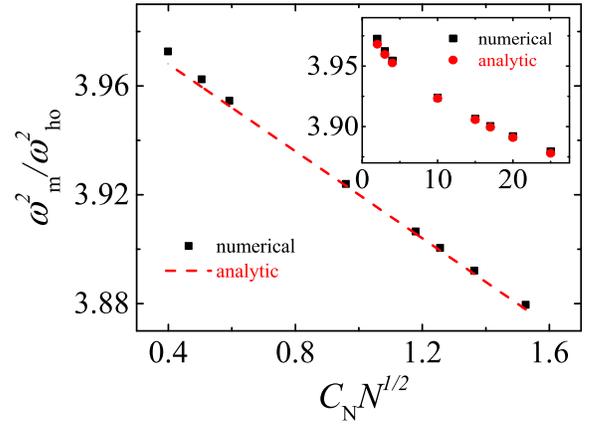}
\caption{\label{fig:4} (Color online) The ratio of the squared breathing mode frequency $\omega_{\rm{m}}^2/\omega_{\rm{ho}}^2$ as a function of $C_{\rm{_N}}\sqrt{N}$ (black squares), compared with the analytic prediction (Eq.~\eqref{eq:neffect_tg}) that is shown by the red dashed line. The inset shows the same ratio as a function of the particle number $N$. Here, we take an interaction strength $g_{\rm{_{1D}}}=10^2$.}
\end{figure}

In Fig.~\ref{fig:4}, we verify this linear behaviour by taking a fixed $g_{\rm{_{1D}}}=10^2$ in the Tonks-Girardeau regime, for which the effective interaction parameter $\gamma_{\rm{_{eff}}}$ varies from $44$ to $157$. 

\subsubsection{Finite temperature effect in the weak coupling limit\label{sec:finitet}}
Here we consider the finite temperature effect in the weakly interacting limit, by calculating collective mode frequencies using the weak-coupling HFBP theory. It is known that at sufficiently low temperature, this effect is small since the thermal fraction of the system is negligible.For example, Debbie Jin's group has shown that the measured collective oscillating frequencies of a 3D Bose gas at temperature $T<0.48T^0_c$ have a good agreement with the theoretical predictions at $T=0$~\cite{PhysRevLett.78.764}. To emphasize the effect of finite temperature on the collective mode, we consider here $T>0.48T^0_c$.

\begin{figure}[t]
\centering
\includegraphics[scale=0.30]{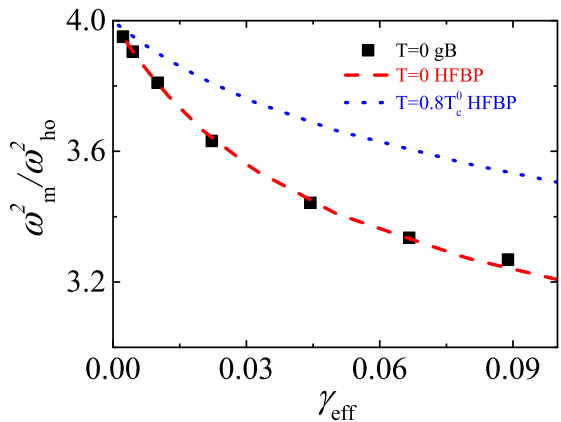}
\caption{\label{fig:5} (Color online) The ratio of the squared breathing mode frequency $\omega_{\rm{m}}^2/\omega_{\rm{ho}}^2$ at $N=25$ in the weakly-interacting regime. We have shown the results at $T=0$ predicted by the generalized Bogoliubov theory (black squares) and by the HFBP theory (red dashed line) and the results at $T=0.8T_{\rm{c}}^0$ given by the HFBP theory (blue dotted line). }
\end{figure}

In Fig.~\ref{fig:5}, we compare the ratios of the squared breathing mode frequency at $T=0$ and $T=0.8T^0_c$. The mode frequency becomes larger at finite temperature. At sufficiently large temperature, actually we anticipate that the ratio approaches the ideal gas limit, i.e., $\omega_{\rm{m}}^2/\omega_{\rm{ho}}^2 = 4$. We also compare the zero temperature ratios, predicted by the generalized Bogoliubov theory (symbols, indicated as gB in the figure) and the HFBP theory (dashed line). There is a good agreement, as anticipated.

\subsubsection{Comparisons with the experiment and previous theory \label{sec:comparison}}
\begin{figure}[t]
\centering
\includegraphics[scale=0.35]{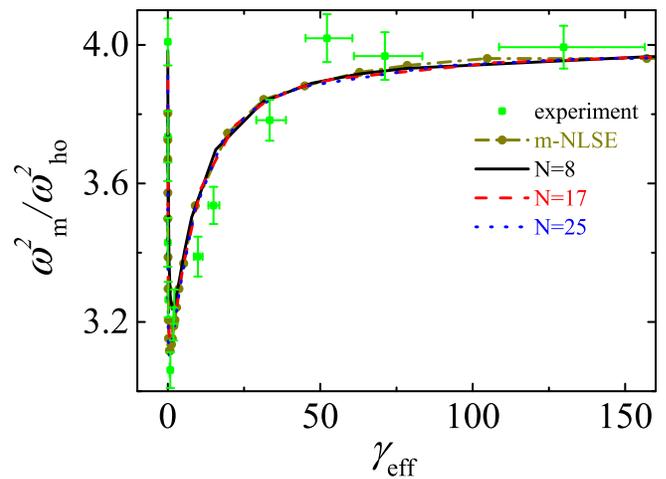}
\caption{\label{fig:6} (Color online) The ratio of the squared breathing mode frequency $\omega_{\rm{m}}^2/\omega_{\rm{ho}}^2$ as a function of the effective interaction parameter $\gamma_{\rm{_{eff}}}$. $\gamma_{\rm{_{eff}}}$ covers all interaction regimes and varies from $2.2\times10^{-3}$ to $2.9\times10^2$. We consider three particle numbers: $N=8$ (black solid line), $N=17$ (red dashed line) and $N=25$ (blue dotted line). We have compared our result with a previous theoretical prediction obtained by using time-dependent modified nonlinear Schr\"{o}dinger equation (m-NLSE) (yellow dot-dashed line)~\cite{2014arXiv1412.6855C} and the experimental data (green squares with error bars)~\cite{Haller04092009}.}
\end{figure}

\begin{figure}[t]
\centering
\includegraphics[scale=0.35]{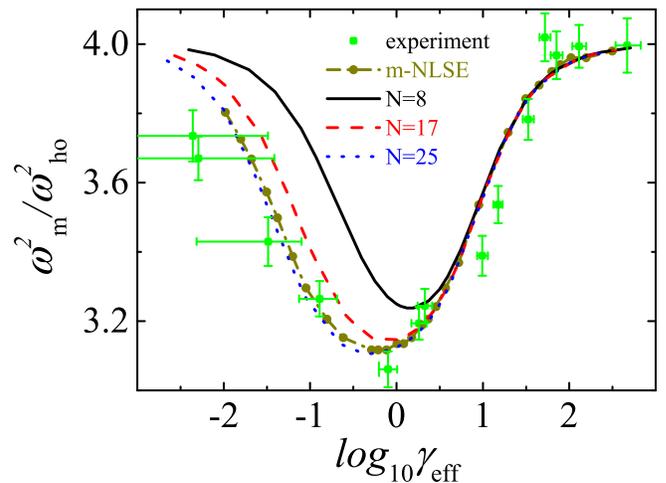}
\caption{\label{fig:7} (Color online) The ratio of the squared breathing mode frequency $\omega_{\rm{m}}^2/\omega_{\rm{ho}}^2$ as a function of $\log_{\rm{10}}\gamma_{_{\rm{eff}}}$. The plot is the same as Fig.~\ref{fig:6}, but is shown here as a function of the interaction parameter in a logarithmic scale, in order to emphasize the particle number dependence in the non-interacting limit. }
\end{figure}

We now address the breathing mode frequency in all interaction regimes, emphasizing its dependence on finite particle number and nonzero temperature.  We vary the effective interaction parameter $\gamma_{\rm{_{eff}}}$ from $2.2\times10^{-3}$ to $2.9\times10^2$, and thus cover all the regimes from the non-interacting limit, the mean-field regime to the Tonks-Girardeau limit. The results are presented as a function of the interaction parameter shown in the linear (Fig.~\ref{fig:6}) and logarithmic scales (Fig.~\ref{fig:7}), in comparison with the experimental data~\cite{Haller04092009} and a previous theoretical prediction~\cite{2014arXiv1412.6855C}.

In general, in these figures the squared frequency ratio $\omega_{\rm{m}}^2/\omega_{\rm{ho}}^2$ of the breathing mode decreases from $4$ in the non-interacting limit to $3$ in the weakly-interacting mean-field regime, and then increase back to $4$ again in the strongly-interacting Tonks-Girardeau regime.

In greater detail, the previous theoretical work (see the results indicated as m-NLSE in the figures) considered a particle number $N=25$~\cite{2014arXiv1412.6855C}. Here we have performed numerical calculations with the same number of particles. We have also considered other two sets of particle number, $N=8$ and $N=17$, since in the experiment the range of the particle number $N$ is $8\sim25$~\cite{Haller04092009}. For the case with a particle number $N=25$, our results agree very well with the m-NLSE predictions. The good agreement is easy to understand, as both theories start from the same generalized Gross-Pitaevskii equation. The different numerical treatments, i.e., the time-dependent simulations in Ref.~\cite{2014arXiv1412.6855C} and our solution of the generalized Bogoliubov equations, only lead to a negligible difference. By further comparing both theoretical predictions at $N=25$ with the experimental data, we find a good agreement in the mean-field and Tonks-Girardeau regimes, where the breathing mode frequency essentially does not depend on the particle number. However, near the non-interacting limit, the discrepancy between experiment and theory becomes evident: the experimental data lie systematically below the theory curves. In this limit, the particle number dependence of the breathing mode frequency is significant.

The particle number dependence is particularly clearly seen in Fig.~\ref{fig:7}. The decreasing of the particle number $N$ from $25$ to $8$ increases the ratio of the squared breathing mode frequency. Thus, taking into account the possibility of a smaller particle number (i.e., $N<25$) in the real experiment will even enlarge the discrepancy between experiment and theory. On the other hand, this discrepancy cannot be resolved as a finite temperature effect, as in the previous subsection we have already examined that a nonzero temperature generally leads to a larger mode frequency. 

\begin{figure}[t]
\centering
\includegraphics[scale=0.35]{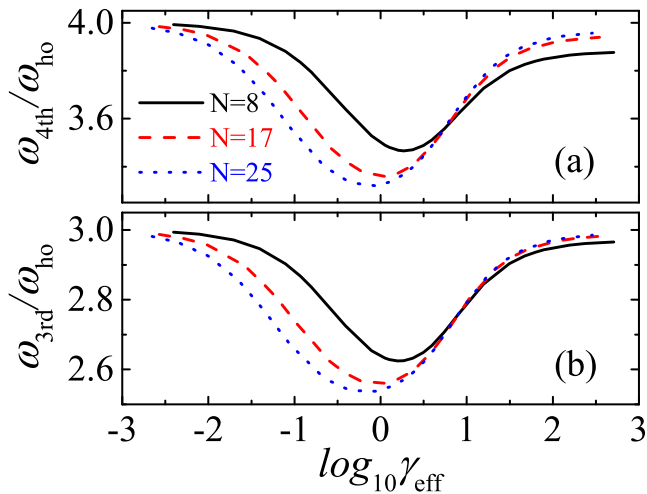}
\caption{\label{fig:8} (Color online) The frequency of higher order compressional modes, $\omega_{\rm{3rd}}/\omega_{\rm{ho}}$ (a) and $\omega_{\rm{4th}}/\omega_{\rm{ho}}$ (b), as a function of $\log_{10}\gamma_{\rm{_{eff}}}$, at different particle numbers: $N=8$ (black solid line), $N=17$ (red dashed line) and $N=25$ (blue dotted line).}
\end{figure}

\subsection{Higher order collective modes}
One of the advantages of our generalized Bogoliubov theory is that we can directly obtain higher order collective mode frequencies from numerical calculations. In Fig.~\ref{fig:8}, we present the mode frequency of the 3-rd (lower panel) and 4-th modes (upper panel) for all interaction regimes at $T=0$, as a function of the effective interaction parameter $\log_{10}\gamma_{\rm{_{eff}}}$. Similarly, the frequencies of higher modes exhibit the same reentrant behavior as the breathing mode frequency. 

Three analytic results can be used to understand the reentrant behavior. In the non-interacting limit, the mode frequency of the $n$-th mode is simply $n\omega_{\rm{ho}}$. In the mean-field regime with sufficiently large number of particles, the collective mode frequency can be analytically determined from a hydrodynamic theory, which predicts $\omega_n=\sqrt{n(n+1)/2}\omega_{\rm{ho}}$~\cite{PhysRevA.66.043610,PhysRevA.68.043610}. Therefore, we have $\omega_{_3}\sim2.45\omega_{\rm{ho}}$ and $\omega_{_4}\sim3.16\omega_{\rm{ho}}$ if $N\rightarrow\infty$. With increasing number of particles, the minimum mode frequencies shown in Fig.~\ref{fig:8} seem to approach these limiting values. Finally, in the Tonks-Girardeau limit, the mode frequency 
of the $n$-th mode again approach $n\omega_{\rm{ho}}$, due to the effective fermionization of the system~\cite{PhysRevA.64.033605}.

It is interesting to note that for high-lying collective modes, the effect of a finite particle number also becomes significant in the Tonks-Girardeau regime as well as in the mean-field regime. This is particularly evident for the 4-th mode, as shown in the upper panel of Fig.~\ref{fig:8}.

\section{Summary and Outlook\label{sec:summary}}
In conclusions, we have studied collective modes of a 1D harmonically trapped Bose gas, by developing two numerical approaches: (a) a generalized Bogoliubov theory at zero temperature following the idea of the density-functional approach, where we have taken the local chemical potential calculated from the Lieb-Liniger model as the exchange energy; (b) a conventional Hartree-Fock-Bogoliubov theory with Popov approximation at finite temperature, where the thermal density is included to make the theory self-consistent. By using these two approaches and by emphasizing the effect of finite particle number and nonzero temperature, we have presented a systematic investigation of the breathing mode frequency in all interaction regimes and have explained the reentrant behaviour of the mode frequency, which varies from 4 in the non-interacting limit, to 3 in the mean-field regime and then back to 4 again in the Tonks-Girardeau limit. The frequency of higher order collective modes exhibits a similar reentrant behaviour. 

We have compared our result with the recent experimental measurement and a previous theoretical prediction. While our result agrees well with the previous theoretical prediction, we have found that both theories cannot explain the measured mode frequency in the non-interacting limit. The discrepancy between experiment and theory becomes even larger when we take a small number of particles or a non-zero temperature. Therefore, we believe more theoretical investigations should be committed in the future in order to fully solve the discrepancy. Those works could focus on the issues such as the inter-tube tunneling and the confinement-induced three-body interparticle interaction. 

\begin{acknowledgments}
We thank H.-C. N\"{a}gerl, M. J. Mark, M. Olshanii, and S. Choi for providing us their data. This work was supported by the ARC Discovery Projects (Grant Nos. DE150101636, FT130100815 and DP140103231).
\end{acknowledgments}

\nocite{*}
\bibliography{bosegas1d}

\end{document}